\documentclass{article}
\usepackage{spconf,amsmath,graphicx}
\usepackage[flushleft]{threeparttable}
\usepackage{cite}
\usepackage{booktabs, multirow}
\usepackage{enumitem} %缩进itemize
\usepackage{amssymb}
\usepackage{hyperref}
\usepackage[table]{xcolor}
\usepackage{pgf}

\definecolor{heatblue}{RGB}{65,115,200}
\definecolor{heatred}{RGB}{205,92,92}
\newcommand{\heatpos}[1]{%
  \pgfmathtruncatemacro{\shadevalue}
    {min(54,14+40*sqrt((#1)/10))}%
  \edef\heatcolor{heatblue!\shadevalue}%
  \expandafter\cellcolor\expandafter{\heatcolor}%
  $+$#1%
}
\newcommand{\heatneg}[1]{%
  \pgfmathtruncatemacro{\shadevalue}
    {min(54,14+40*sqrt((#1)/10))}%
  \edef\heatcolor{heatred!\shadevalue}%
  \expandafter\cellcolor\expandafter{\heatcolor}%
  $-$#1%
}
\newcommand{\heatposbest}[1]{%
  \pgfmathtruncatemacro{\shadevalue}
    {min(54,14+40*sqrt((#1)/10))}%
  \edef\heatcolor{heatblue!\shadevalue}%
  \expandafter\cellcolor\expandafter{\heatcolor}%
  $\mathbf{+}$\textbf{#1}%
}
\newcommand{\heatnegbest}[1]{%
  \pgfmathtruncatemacro{\shadevalue}
    {min(54,14+40*sqrt((#1)/10))}%
  \edef\heatcolor{heatred!\shadevalue}%
  \expandafter\cellcolor\expandafter{\heatcolor}%
  $\mathbf{-}$\textbf{#1}%
}

\title{CASM: Context-Aware Semi-Markov Post-Processor for Beat Tracking}
\name{Zhanhong He$^{1}$, Hanyu Meng$^{2}$, Yaolong Ju$^{3,}$\sthanks{Corresponding author, juyaolong@gbu.edu.cn.}}
\address{
    $^{1}$The University of Western Australia, Perth, Australia\\
    $^{2}$The University of New South Wales, Sydney, Australia\\
    $^{3}$Great Bay University, Dongguan, China
}

\begin{document}
\ninept
\maketitle

\begin{abstract}
In music beat tracking, model-predicted activations must be decoded into a discrete, musically coherent event sequence. Direct peak picking closely follows local evidence but can retain spurious peaks or miss weak beats. Dynamic Bayesian networks (DBNs), a widely used structured post-processor, improve sequence consistency under predefined global tempo, meter, and transition constraints, but their behavior can depend strongly on these settings. We introduce CASM, a context-aware semi-Markov decoder that instead conditions its temporal constraint on local activation evidence. CASM also accounts for ambiguity among competing periodic interpretations, including half- and double-tempo alternatives. Deterministic safeguards prevent implausible outputs and preserve beat--downbeat consistency. Applied to fixed activations from three neural beat trackers (BeatThis, MSCNN, and TCN), CASM improves temporal continuity while preserving event-level F1 across the GTZAN and SMC datasets, without backbone retraining or dataset-specific retuning. Further analysis shows that CASM is less sensitive than the DBN baseline to the composition of the calibration data.\footnote{Code and demo are available at \url{https://zhanh-he.github.io/casm-beat-tracking}}
\end{abstract}


\section{Introduction}
\label{sec:intro}
Beats give music its temporal scaffold, so reliable beat tracking is crucial for music production, analysis, synchronization, and interactive systems~\cite{purwins2019deep}. More recently, beat tracking has become a key component of multimodal AI-generated content (AIGC), where the estimated beats serve as timing anchors for aligning generated content such as dance~\cite{huang2024beatit}, video~\cite{yu2023loris, tong2026vem}, and music~\cite{lan2024musicongen, chen2024musicldm}. Most music beat trackers operate in two stages: a network maps audio to framewise beat activations, and a post-processor converts these activations into discrete events. Despite recent attempts to remove the post-processor~\cite{chen2022postprocessingfree, foscarin2024beat, bolt2026reformulated}, many contemporary systems \cite{deng2024efficient, ru2025hingenet, ru2025beatfm, ru2026beatmamba, zhang2025beatkan} still rely on it to obtain musically coherent estimates. No matter how accurate the neural activations become, direct peak picking makes decisions locally without considering the periodicity of beat sequences. Post-processing addresses this limitation by combining local activation evidence with the temporal regularity of music.

For post-processors, the key design question is how to balance temporal regularity against local activation evidence. A weak regularity constraint preserves the network's predictions but may leave missed or spurious beats uncorrected. An overly strong constraint can instead suppress valid peaks under expressive timing or rapid local tempo variation, which is common in classical music performances~\cite{Schreiber2020LocalTempo}.

Many existing post-processors tilt this balance toward predefined temporal constraints. Early dynamic programming approach~\cite{ellis2007beattracking} assumes a single global tempo per track and penalizes inter-beat intervals (IBIs) that deviate from it. Conditional random fields (CRFs)~\cite{fillon2015crf, fuentes2019skipchain} learn sequence preferences that permit a range of local tempo variation while discouraging abrupt IBI changes. The dominant post-processor is the dynamic Bayesian network (DBN)~\cite{krebs2015dbn, bock2016joint}, which operates under pre-specified tempo ranges, meter states, and tempo transition probabilities. This design works well when these assumptions match the recording, but may require per-track or corpus-specific retuning when they do not~\cite{ahn2026smcblindspot}.

The recently proposed PLPDP~\cite{chiu2023localperiodicity} shifts this balance to local evidence-conditioned temporal constraint. It derives a time-varying IBI and confidence from the predominant local pulse (PLP), which set the preferred period and how strongly it is enforced. However, because this confidence is derived from PLP peak heights, it reflects how strong the selected period is but not whether a competing period is nearly as plausible. A strong period may still be nearly tied with a metrical alternative, such as half or double tempo, and enforcing it strongly then risks committing to the wrong metrical level.
 
We introduce CASM, an offline context-aware semi-Markov post-processor. While semi-Markov decoding has shown leading performance and efficiency in real-time beat tracking \cite{heydari2022semimarkov}, CASM operates offline on the complete activation sequence (with access to future context), so its performance is not directly comparable with that of causal online systems \cite{meier2024realtime, heydari2024beatnetplus}. The core philosophy is simple. Whereas PLPDP measures periodicity strength, CASM additionally accounts for periodicity ambiguity. It constructs a sparse path over activation-supported peaks and enforces strict temporal regularity when periodic cues are clear, but relaxes this constraint to let neural activations dominate when competing periods create ambiguity. A beat-count safeguard and a beat-synchronous meter decoder bound this behavior and keep the decoding deterministic. CASM offers three core advantages:
\begin{itemize}[leftmargin=*, nosep]
\item \textbf{Backbone-agnostic and transferable.} CASM is a plug-in post-processor to cooperate with existing beat trackers and requires no backbone retraining. With a single fixed configuration, it improves every metric in our controlled comparison across BeatThis \cite{foscarin2024beat}, MSCNN \cite{he2026dynest}, and TCN \cite{bock2020deconstruct} on both SMC \cite{smc} and GTZAN \cite{gtzan} datasets.
\item \textbf{Low tuning burden.} Like PLPDP, CASM derives the preferred interval and constraint strength from each activation sequence, avoiding per-track or per-dataset BPM and meter retuning.
\item \textbf{Open and reproducible.} We release CASM as an open-source Python package with the configuration used in our experiments and a \textit{madmom}-like activation-to-event interface.
\end{itemize}

\begin{figure}[htbp]
\centering
\includegraphics[width=\columnwidth]{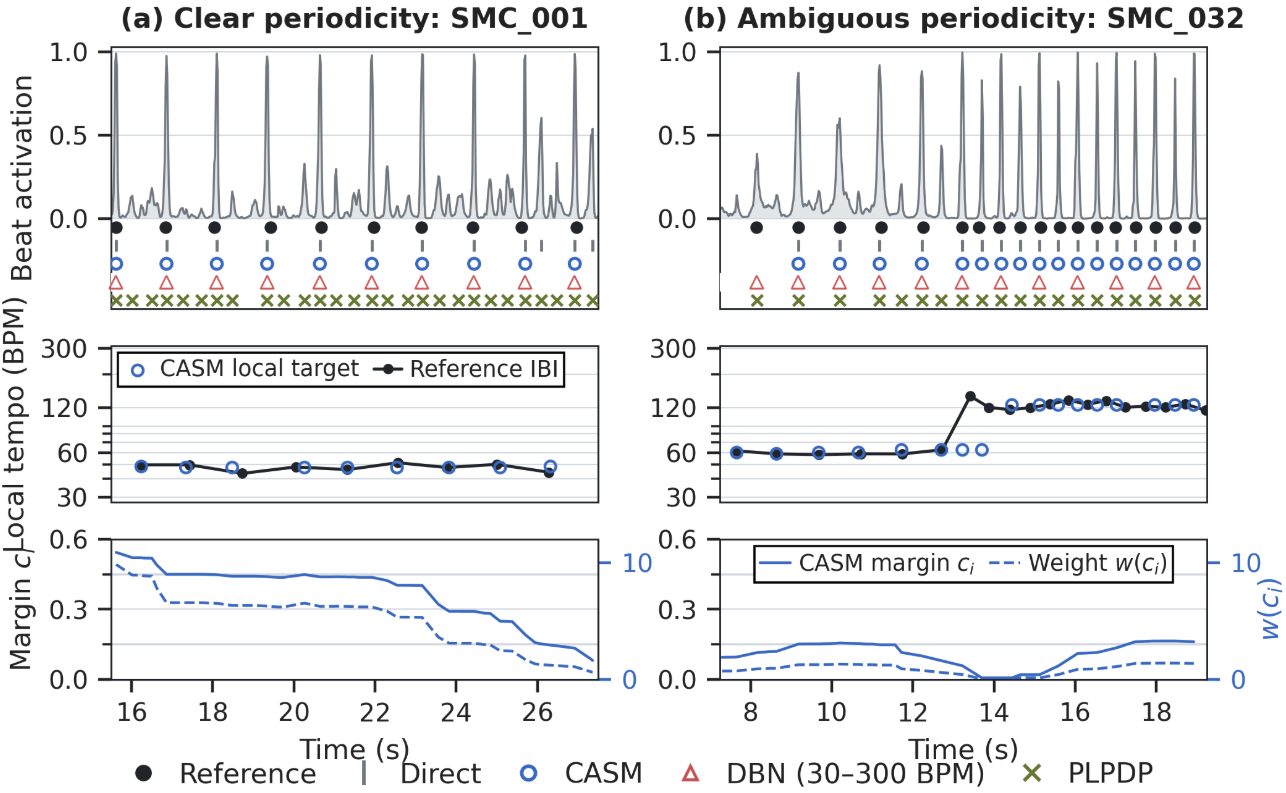}
\par\vspace{-2mm}
\caption{CASM behavior on BeatThis activations. (a) CASM enforces stronger temporal regularization when periodicity is clear, but (b) closely follows direct peak-picking outputs when periodicity is ambiguous. $w(c_i)$ denotes the strength of temporal regularization.}
\label{fig:overview}
\end{figure}

\section{Methodology}
\label{sec:method}

\subsection{CASM overview}
\label{sec:overview}
Figure~\ref{fig:overview} illustrates the adaptive role of CASM. When local periodicity is clear, CASM strengthens temporal regularity. When it is ambiguous, CASM relaxes the constraint and follows the neural activations more closely. CASM is an offline, model-agnostic post-processor operating on the beat and downbeat activations of a frozen neural tracker. It selects beats among activation peaks with a semi-Markov path that balances activation evidence and interval regularity (Sec.~\ref{sec:path}). At each candidate peak, CASM estimates the preferred local period and its margin over the strongest competitor (Sec.~\ref{sec:margin}). The period sets the preferred inter-beat interval, while the margin controls how strongly it is enforced (Sec.~\ref{sec:cost}). Decoding, safeguards, and downbeats complete the method (Sec.~\ref{sec:safeguard}).

\subsection{Semi-Markov beat path}
\label{sec:path}
CASM keeps every activation peak above a small threshold, including weak ones, as a candidate beat. Each candidate $i$ receives an evidence score $e_i$, its beat logit scaled and clipped to a bounded range, so that $e_i>0$ exactly when the activation exceeds 0.5. Two candidates can be consecutive beats only if the time $\Delta_{ij}$ between them lies within the interval support of 0.2--2\,s (30--300\,BPM). CASM selects the candidate sequence $\pi$ that maximizes
\begin{equation}
\mathcal S(\pi)=\sum_{i\in\pi}e_i-\sum_{(i,j)\in\pi}D_{ij},
\label{eq:casm-objective}
\end{equation}
where $(i,j)$ runs over consecutive beats of $\pi$ and $D_{ij}$ is the duration cost of their interval. The two terms respectively reward strong beat evidence and penalize rhythmically inconsistent intervals.

This is the sense in which CASM is semi-Markov: each transition jumps from one beat directly to the next and scores the elapsed interval as a whole. A frame-level DBN tracks tempo and beat position as hidden states; CASM instead reads the preferred interval from the local context, so its decoding state is simply the last selected beat and no tempo states are needed. The dynamic programs of Ellis~\cite{ellis2007beattracking} and PLPDP~\cite{chiu2023localperiodicity} also score beat-to-beat intervals, but over every frame rather than only activation peaks. The scores are hand-specified, so CASM is a deterministic decoder rather than a generative hidden semi-Markov model or a learned semi-CRF.

\subsection{Local period and period margin}
\label{sec:margin}
To estimate the local period and its ambiguity, CASM compares the beat activation in an 8-s window centered on each candidate $i$ with lagged copies of itself, using normalized autocorrelation over lags of 0.2--2\,s. Because autocorrelation also peaks at multiples of the true period, a mild prior favors shorter lags. The best lag gives the local period $\tau_i$ with score $q_i^\star$, and $q_i^{\mathrm{alt}}$ is the best score among lags more than two frames away from it. The period margin
\begin{equation}
c_i=\frac{q_i^\star-q_i^{\mathrm{alt}}}{q_i^\star}
\label{eq:casm-margin}
\end{equation}
lies between 0 and 1. A large $c_i$ indicates a clearly dominant period, whereas a small value indicates ambiguity from nearby or half- or double-time alternatives. It is a deterministic separation margin rather than a calibrated probability. Both $\tau_i$ and $c_i$ are median-filtered over five consecutive candidates.

\subsection{Margin-conditioned duration cost}
\label{sec:cost}
Each interval $(i,j)$ uses the context of both endpoints through geometric means, $\tau_{ij}=\sqrt{\tau_i\tau_j}$ and $c_{ij}=\sqrt{c_ic_j}$, so it is treated as unambiguous only when both of its ends are. Its duration cost is
\begin{equation}
D_{ij}=c_{ij}\,\frac{\lambda}{2\,\sigma(c_{ij})^2}\,\log^2\!\frac{\Delta_{ij}}{\tau_{ij}},
\quad
\sigma(c)=\sigma_0+(1-c)\,\sigma_{\mathrm u},
\label{eq:casm-duration}
\end{equation}
a Gaussian penalty on the log-interval that is zero at $\Delta_{ij}=\tau_{ij}$ and treats doubled and halved intervals alike. Here, $\lambda>0$ controls the overall penalty strength, $\sigma_0>0$ sets the minimum timing tolerance, and $\sigma_{\mathrm u}\geq0$ controls relaxation under ambiguity. The margin thus plays two roles: the factor $c_{ij}$ sets the penalty weight and $\sigma(c_{ij})$ its tolerance. As $c_{ij}$ falls, the weight drops and the tolerance widens, so the cost vanishes at $c_{ij}=0$ and stiffens rapidly toward $c_{ij}=1$.

This single cost lets CASM add and remove beats relative to Direct: a weak candidate inside a doubled interval is selected if the two regular halves save more cost than its low evidence loses, and a strong off-grid peak is skipped if its irregular intervals cost more than its evidence gains. A large $c_{ij}$ tilts these trade-offs toward a regular path, whereas a small one lets the activations decide, moving the path toward Direct. CASM can therefore recover weak but rhythmically supported beats without forcing an unreliable tempo grid.

\subsection{Decoding, safeguards, and downbeats}
\label{sec:safeguard}
A Viterbi-style dynamic program maximizes Eq.~\eqref{eq:casm-objective} and backtracks the best beat path, allowing a restart after an unfavorable prefix. Because the centered window and the backtracking use future activations, CASM is an offline decoder. As a count safeguard, CASM also forms the Direct path, i.e., the minimal peak picking of BeatThis~\cite{foscarin2024beat} with seven-frame max pooling, a zero-logit threshold, and one-frame deduplication. If the CASM path has fewer than 0.85 or more than 1.8 times as many beats as Direct ($\lvert\pi_{\mathrm{CASM}}\rvert/
 \lvert\pi_{\mathrm{direct}}\rvert
 \notin[0.85,1.8]$), indicating an implausibly sparse or dense path, the Direct path is returned. This rule is deterministic and requires no learned router.

Downbeats are decoded on the selected beat grid using a second dynamic program that allows bars of 2--7 beats and discourages unnecessary meter changes. If its output agrees poorly with the Direct downbeat peaks snapped to the same grid (F-measure below 0.6 at $\pm$70\,ms), the snapped result is used, ensuring that every downbeat coincides with an output beat. All scalar parameters are calibrated once on the development folds (Sec.~\ref{sec:calib}) and then frozen across tracks, datasets, and backbone models.

\section{Experiments}
\label{sec:exp}
\subsection{Evaluation protocol}
\textbf{Datasets and splits.} We use the publicly released data from BeatThis \cite{foscarin2024beat}, which contains 4,556 tracks from 18 beat-tracking datasets, including SMC, Hainsworth, Ballroom, Beatles, and others. This data collection provides preprocessed log-Mel spectrograms computed from 22.05\,kHz mono audio at 50 fps, together with a fixed eight-fold partition to conduct cross-validation experiments. For each fold, the corresponding backbone checkpoint is trained on the remaining seven folds, ensuring that evaluated tracks remain unseen during training. Among these datasets, SMC provides beat but not downbeat annotations, and is widely regarded as the most challenging among them. Additionally, GTZAN provides 993 tracks as the held-out evaluation set, which is excluded from the cross-validation pool.

\textbf{Backbones.} We select three models whose implementations are publicly available and readily reproducible. For BeatThis, we use the original code and pretrained checkpoints \cite{foscarin2024beat}. MSCNN is adapted from the multi-scale, multi-task convolutional network of \cite{he2026dynest}, which was designed for 60\,s Bark-scale specific-loudness features. We modify it to accept the 30s the unified log-Mel input and remove the task heads unrelated to beat and downbeat tracking. TCN was introduced by \cite{bock2020deconstruct}, and we use the re-implementation distributed with \cite{he2026dynest}, which has been adapted from its original 5\,s input to the same 30\,s log-Mel input, with the tempo task head removed. All resulting backbones are trained on the same data and frozen before decoding performance comparison, ensuring that every post-processor receives exactly the same 50\,Hz framewise beat/downbeat activations from a given backbone.

\textbf{Evaluation.} We conduct the same eight-fold cross-validation as in \cite{foscarin2024beat}, where backbones are each evaluated on the held-out GTZAN and their corresponding out-of-fold SMC partitions. We use \texttt{mir\_eval} \cite{raffel2014mireval} to compute F1 score with a $\pm70$\,ms tolerance, CMLt, and AMLt. CMLt measures continuity at the correct metrical level, whereas AMLt additionally accepts standard off-beat and half- or double-level interpretations.

\subsection{Post-processors and tempo support}
Since our task is offline, all post-processors can use future context. We compare CASM with direct peak picking and three offline post-processors including DBN~\cite{krebs2015dbn}, PLPDP~\cite{chiu2023localperiodicity}, and CRF beat detection~\cite{korzeniowski2014crf}. Direct peak-picking has been defined in Sec.~\ref{sec:safeguard}, and CASM uses a 30--300-BPM range by default. PLPDP and CRF are run with their released/default parameters. For methods without a joint beat/downbeat decoder, downbeats are decoded from the downbeat activation stream and snapped to the corresponding beat grid.

DBN receives separate treatment because its behavior is particularly sensitive to global tempo support. We use the standard joint \texttt{madmom} configuration \cite{krebs2015dbn}: meters $\{3,4\}$, transition weight 100, observation weight 16, threshold 0.05, and the default 55--215-BPM range. Since many SMC excerpts have tempi below 55 BPM \cite{ahn2026smcblindspot}, Table~\ref{tab:postprocessor_ablation} also explores a 30--300-BPM DBN variant in which only the tempo range is changed. For a matched diagnostic, we additionally report CASM with its BPM range restricted to 55--215, with all other settings unchanged (Table~\ref{tab:casm_ablation}).

\subsection{CASM calibration}
\label{sec:calib}
CASM has no learned weights or random seed. Calibration is a simple grid search that keeps the configuration with the highest beat F1 on the calibration folds. The selected values depend on the calibration data used in each experiment, so we do not list them here. The stable released parameters and the full tuning history are available in our public GitHub repository. The configuration used in all main tables is calibrated once on SMC folds 1--7 (7F) and then frozen, so the same settings are used for every track, backbone, and dataset. SMC fold 0 is held out from decoder calibration and serves, together with GTZAN, as unseen evaluation data in Fig.~\ref{fig:calibration-scale}.

To test sensitivity to calibration data, we calibrate on every subset of 1, 2, 4, or 7 folds from folds 1--7, denoted as 1F, 2F, 4F, and 7F, respectively. These scales yield 7, 21, 35, and 1 combinations, giving 64 calibration sets per decoder with no hand-picked combination. CASM and DBN are calibrated on each set using the same selection rule. For DBN, the grid covers the tempo range, the transition weight, which controls how strongly tempo changes are penalized, the observation weight, and the threshold. Every selected configuration is then evaluated on SMC fold 0, which tests transfer to unseen excerpts from the same collection, and on GTZAN using one fixed backbone seed, which tests transfer to a different dataset.

% \vspace{-3pt}

\section{Results}
\label{sec:results}

\subsection{Comparison of post-processing methods}
\label{sec:post_comp}
Table~\ref{tab:postprocessor_ablation} isolates post-processor performance by holding the backbone activations fixed. CASM yields numerical improvements over the direct peak-picking baseline across every reported metric for all three backbones. While the 30--300,BPM DBN can yield large gains in continuity (particularly for GTZAN downbeats), it consistently degrades F1 accuracy. Furthermore, DBN performance is sensitive to tempo support: restricting its range to the default 55--215,BPM range improves some GTZAN metrics but causes large performance drops on SMC. Other alternatives such as PLPDP and CRF provide isolated benefits but fail to improve metrics consistently. Overall, CASM serves as a robust, plug-and-play upgrade that balances accuracy and continuity without requiring backbone-specific tuning.

\vspace{-3pt}

\begin{table}[!h]
\centering
\caption{Post-processing comparison on fixed backbone activations. Backbone-only rows use direct peak picking. Post-processors use 30--300\,BPM unless marked $^{[55,215]}$, the default DBN range which optimized for GTZAN-like music.}
\label{tab:postprocessor_ablation}
\begingroup
\fontsize{8pt}{8.5pt}\selectfont
\setlength{\tabcolsep}{0.3pt}
\renewcommand{\arraystretch}{0.92}
\resizebox{\columnwidth}{!}{%
\begin{tabular}{l|rrr|rrr|rrr}
\toprule
\multicolumn{1}{c|}{\multirow{2}{*}{\textbf{Method}}}
& \multicolumn{6}{c|}{\textbf{GTZAN (Beat / Downbeat)}}
& \multicolumn{3}{c}{\textbf{SMC (Beat)}} \\
\cmidrule(lr){2-7}
\cmidrule(l){8-10}
& F1 & CMLt & AMLt
& F1 & CMLt & AMLt
& F1 & CMLt & AMLt \\
\midrule
\textbf{BeatThis}\cite{foscarin2024beat}
& 89.1 & 79.8 & 89.8
& 78.3 & 67.2 & 79.1
& 62.7 & 51.4 & 61.0 \\
w. CASM
& \heatposbest{0.3} & \heatposbest{0.8} & \heatpos{1.2}
& \heatposbest{0.7} & \heatpos{3.6} & \heatpos{8.0}
& \heatposbest{0.3} & \heatposbest{2.4} & \heatposbest{3.6} \\
% w. CASM$^{\mathrm{[dbn]}}$
% & \heatpos{0.1} & \heatpos{0.5} & \heatpos{0.6}
% & \heatpos{0.5} & \heatpos{2.9} & \heatpos{5.8}
% & \heatpos{0.1} & \heatpos{0.4} & \heatpos{2.0} \\
w. DBN \cite{krebs2015dbn}
& \heatneg{0.5} & \heatneg{0.9} & \heatpos{0.2}
& \heatneg{1.5} & \heatneg{0.6} & \heatpos{1.1}
& \heatneg{1.6} & \heatpos{0.4} & \heatpos{3.5} \\
w. DBN$^{[55,215]}$
& \heatneg{1.0} & \heatpos{0.7} & \heatpos{1.3}
& \heatneg{0.9} & \heatposbest{6.2} & \heatposbest{8.7}
& \heatneg{5.2} & \heatneg{3.9} & \heatpos{1.5} \\
w. PLPDP \cite{chiu2023localperiodicity} \space{}
& \heatpos{0.1} & \heatpos{0.3} & \heatpos{0.3}
& \heatpos{0.1} & \heatpos{0.2} & \heatpos{0.2}
& \heatneg{0.9} & \heatneg{3.5} & \heatneg{0.3} \\
w. CRF \cite{korzeniowski2014crf}
& \heatneg{0.6} & \heatneg{0.3} & \heatposbest{1.4}
& \heatneg{1.4} & \heatpos{1.9} & \heatpos{5.5}
& \heatneg{1.2} & \heatpos{1.4} & \heatpos{1.2} \\
\midrule
\textbf{MSCNN}\cite{he2026dynest}
& 86.1 & 73.4 & 79.9
& 70.8 & 53.1 & 63.1
& 57.2 & 42.0 & 48.4 \\
w. CASM
& \heatposbest{0.4} & \heatpos{1.6} & \heatpos{2.4}
& \heatpos{1.5} & \heatpos{12.5} & \heatpos{18.3}
& \heatposbest{0.3} & \heatposbest{3.5} & \heatposbest{6.8} \\
% w. CASM$^{\mathrm{[dbn]}}$
% &&&
% &&&
% &&& \\
w. DBN
& \heatneg{1.1} & \heatpos{0.2} & \heatpos{5.9}
& \heatneg{0.9} & \heatpos{9.5} & \heatpos{14.6}
& \heatneg{0.2} & \heatpos{2.2} & \heatpos{4.1} \\
w. DBN$^{[55,215]}$
& \heatneg{0.6} & \heatposbest{4.0} & \heatposbest{7.2}
& \heatposbest{1.8} & \heatposbest{15.8} & \heatposbest{21.2}
& \heatneg{4.9} & \heatpos{0.3} & \heatpos{3.9} \\
\midrule
\textbf{TCN}\cite{bock2020deconstruct}
& 87.1 & 73.8 & 81.7
& 62.4 & 24.6 & 59.9
& 52.9 & 30.1 & 36.0 \\
w. CASM
& \heatposbest{0.3} & \heatposbest{1.3} & \heatpos{1.6}
& \heatpos{2.1} & \heatpos{9.2} & \heatpos{13.4}
& \heatposbest{1.7} & \heatposbest{4.5} & \heatposbest{8.2} \\
% w. CASM$^{\mathrm{[dbn]}}$
% &&&
% &&&
% &&& \\
w. DBN
& \heatneg{1.9} & \heatneg{0.8} & \heatpos{4.4}
& \heatneg{0.2} & \heatpos{8.9} & \heatpos{9.3}
& \heatneg{0.4} & \heatpos{2.1} & \heatpos{5.9} \\
w. DBN$^{[55,215]}$
& \heatneg{2.6} & \heatneg{0.7} & \heatposbest{5.0}
& \heatposbest{4.9} & \heatposbest{23.2} & \heatposbest{22.9}
& \heatneg{0.9} & \heatneg{7.0} & \heatpos{2.3} \\
\bottomrule
\end{tabular}%
}
\endgroup
\end{table}

\subsection{Ablation of CASM components}
Table~\ref{tab:casm_ablation} details the contribution of each CASM component by changing one design choice at a time. The adaptive margin is the most critical factor. With fixed stiffness, GTZAN gains turn negative and SMC gains shrink sharply. The margin should control both the penalty weight and the tolerance: using it for only one of them lowers F1 on both datasets and removes the GTZAN gains. A single-endpoint context keeps all gains positive but reduces them, especially on GTZAN. Without the count safeguard, F1 falls below the direct peak-picking while continuity improves, so the safeguard is what keeps CASM from trading F1 for continuity. Restricting the tempo range to 55--215 keeps every gain positive but makes them smaller. Unlike DBNs, CASM therefore needs less dataset-specific configuration, since a wider tempo range is better on both datasets.

\vspace{-6pt}

\begin{table}[!h]
\centering
\caption{Ablation of CASM on BeatThis activations. Each row changes one design choice described in the subsection of Sec.~\ref{sec:method}. Values are changes from the direct peak-picking in percentage points.}
\label{tab:casm_ablation}
\begingroup
\fontsize{9pt}{9pt}\selectfont
\setlength{\tabcolsep}{0.5pt}
\renewcommand{\arraystretch}{0.92}
\resizebox{\columnwidth}{!}{%
\begin{tabular}{l|c|rrr|rrr}
\toprule
\multicolumn{1}{c|}{\multirow{2}{*}{\textbf{Variant}}}
& \multirow{2}{*}{\textbf{Sec.}}
& \multicolumn{3}{c|}{\textbf{GTZAN (Beat)}}
& \multicolumn{3}{c}{\textbf{SMC (Beat)}} \\
\cmidrule(lr){3-5}
\cmidrule(l){6-8}
& 
& F1 & CMLt & AMLt
& F1 & CMLt & AMLt \\
\midrule
\textbf{CASM (full)}
& --
& \heatposbest{0.3} & \heatposbest{0.8} & \heatposbest{1.2}
& \heatposbest{0.3} & \heatposbest{2.4} & \heatposbest{3.6} \\
\midrule
55--215-BPM support
& \hyperref[sec:path]{2.2}
& \heatpos{0.1} & \heatpos{0.5} & \heatpos{0.6}
& \heatpos{0.1} & \heatpos{0.4} & \heatpos{2.0} \\
Fixed stiffness (no margin)
& \hyperref[sec:margin]{2.3}
& \heatneg{0.3} & \heatneg{0.8} & \heatneg{0.8}
& \heatpos{0.1} & \heatpos{0.7} & \heatpos{0.9} \\
Margin on weight only
& \hyperref[sec:cost]{2.4}
& \heatneg{0.1} & \heatneg{0.2} & \heatneg{0.2}
& \heatneg{0.5} & \heatpos{0.9} & \heatpos{2.1} \\
Margin on tolerance only
& \hyperref[sec:cost]{2.4}
& \heatneg{0.3} & \heatneg{0.6} & \heatneg{0.6}
& \heatneg{0.5} & \heatpos{0.9} & \heatpos{2.2} \\
Single-endpoint context
& \hyperref[sec:cost]{2.4}
& \heatpos{0.1} & \heatpos{0.3} & \heatpos{0.4}
& \heatpos{0.2} & \heatpos{2.2} & \heatpos{2.6} \\
No count safeguard
& \hyperref[sec:safeguard]{2.5}
& \heatneg{0.1} & \heatpos{0.4} & \heatpos{0.5}
& \heatneg{0.3} & \heatpos{2.5} & \heatpos{2.7} \\
\bottomrule
\end{tabular}%
}
\endgroup
\end{table}

\vspace{-4pt}

\subsection{Comparison with state-of-the-art systems}
Table~\ref{tab:audiodyn_cross_dataset_sota} compares BeatThis with CASM against recent systems, all but one of which rely on DBN post-processing. The exception, BeatThis-MDM~\cite{foscarin2026masked}, uses a learned masked-diffusion decoder trained jointly with the BeatThis network. As it is data-driven, BeatThis-MDM improves on every GTZAN metric, whereas its SMC F1 and CMLt drop below those of BeatThis (direct). A learned decoder may favour their training-like data. CASM, in contrast, needs no retraining and works directly on the activations of any trained backbone, with the same parameters for every backbone. As Section~\ref{sec:post_comp} shows, CASM gives higher beat F1 than DBN and improves all SMC metrics across all tested backbones. DBN-based systems may therefore benefit from CASM. Those systems with large pre-trained frontends keep the best scores \cite{ru2025hingenet, ru2025beatfm}, and replacing DBN with CASM in the latter is a natural next step.

\vspace{-3pt}

\begin{table}[!h]
\centering
\caption{Comparison with recent beat-tracking systems. BeatThis (direct) and CASM results were obtained from our experiments, while all other results are reported from their original papers.}
\label{tab:audiodyn_cross_dataset_sota}
\begingroup
\fontsize{9pt}{10pt}\selectfont
\setlength{\tabcolsep}{0.5pt}
\renewcommand{\arraystretch}{0.95}
\resizebox{\columnwidth}{!}{%
\begin{tabular}{l|ccc|ccc|ccc}
\toprule
\multicolumn{1}{c|}{\multirow{2}{*}{\textbf{Method}}}
& \multicolumn{6}{c|}{\textbf{GTZAN (Beat / Downbeat)}}
& \multicolumn{3}{c}{\textbf{SMC (Beat)}} \\
\cmidrule(lr){2-4}
\cmidrule(lr){5-7}
\cmidrule(l){8-10}
& F1 & CMLt & AMLt
& F1 & CMLt & AMLt
& F1 & CMLt & AMLt \\
\midrule

\textbf{HN-MERT} \cite{ru2025hingenet}
& \textbf{89.7} & 81.4 & \textbf{94.3}
& 77.4 & 71.6 & 87.2
& 61.7 & -- & -- \\

\textbf{HN-MusicFM} 
& 89.2 & 80.9 & 93.7
& \textbf{79.8} & 73.2 & \textbf{89.5}
& 60.6 & -- & -- \\

\textbf{BeatFM-MERT} \cite{ru2025beatfm}
& \underline{89.5} & \underline{82.7} & \underline{93.9}
& 76.7 & 70.2 & 85.4
& 61.3 & -- & -- \\

\textbf{BeatFM-MusicFM}
& 89.1 & 80.6 & 93.5
& \underline{79.6} & \underline{74.4} & \underline{88.7}
& 60.5 & -- & -- \\

\textbf{BeatMamba} \cite{ru2026beatmamba}
& 88.9 & 82.3 & \textbf{94.3}
& -- & -- & --
& 58.7 & 46.8 & 62.4 \\

\textbf{BeatKAN} \cite{zhang2025beatkan}
& 88.2 & 78.1 & 92.3
& -- & -- & --
& 59.8 & 48.4 & 64.0 \\

\midrule
\rowcolor{gray!10}
\multicolumn{10}{c}{%
\textit{Above: with DBN post-processing\quad below: without DBN post-processing.}
} \\
\midrule

\textbf{BeatThis-MDM} \cite{foscarin2026masked}
& \textbf{89.7} & \textbf{82.9} & 92.5
& 79.5 & \textbf{76.4} & 88.5
& 61.0 & 48.9 & \underline{64.3} \\

\textbf{BeatThis} \cite{foscarin2024beat}
& 89.1 & 79.8 & 89.8
& 78.3 & 67.2 & 79.1
& \underline{62.7} & \underline{51.4} & 61.0 \\

\textbf{BeatThis} w. CASM
& 89.4 & 80.6 & 91.0
& 79.0 & 70.8 & 87.1
& \textbf{63.0} & \textbf{53.8} & \textbf{64.6} \\

\bottomrule
\end{tabular}%
}
\endgroup
\end{table}

\subsection{Sensitivity to calibration data}
The sensitivity experiment keeps the evaluation sets fixed (the 0-fold of SMC, and the whole held-out GTZAN) and changes only the calibration data. For each calibration scale $k\in\{1,2,4,7\}$, CASM and DBN are calibrated on every $k$-fold subset of SMC folds 1--7 (Sec.~\ref{sec:exp}) and evaluated on the held-out SMC fold 0 and GTZAN. Thus, the spread at a given scale measures dependence on the chosen folds, while the shift of the distribution centre measures the effect of additional calibration data.

\begin{figure}[!h]
\begin{center}
\includegraphics[width=0.9\columnwidth]{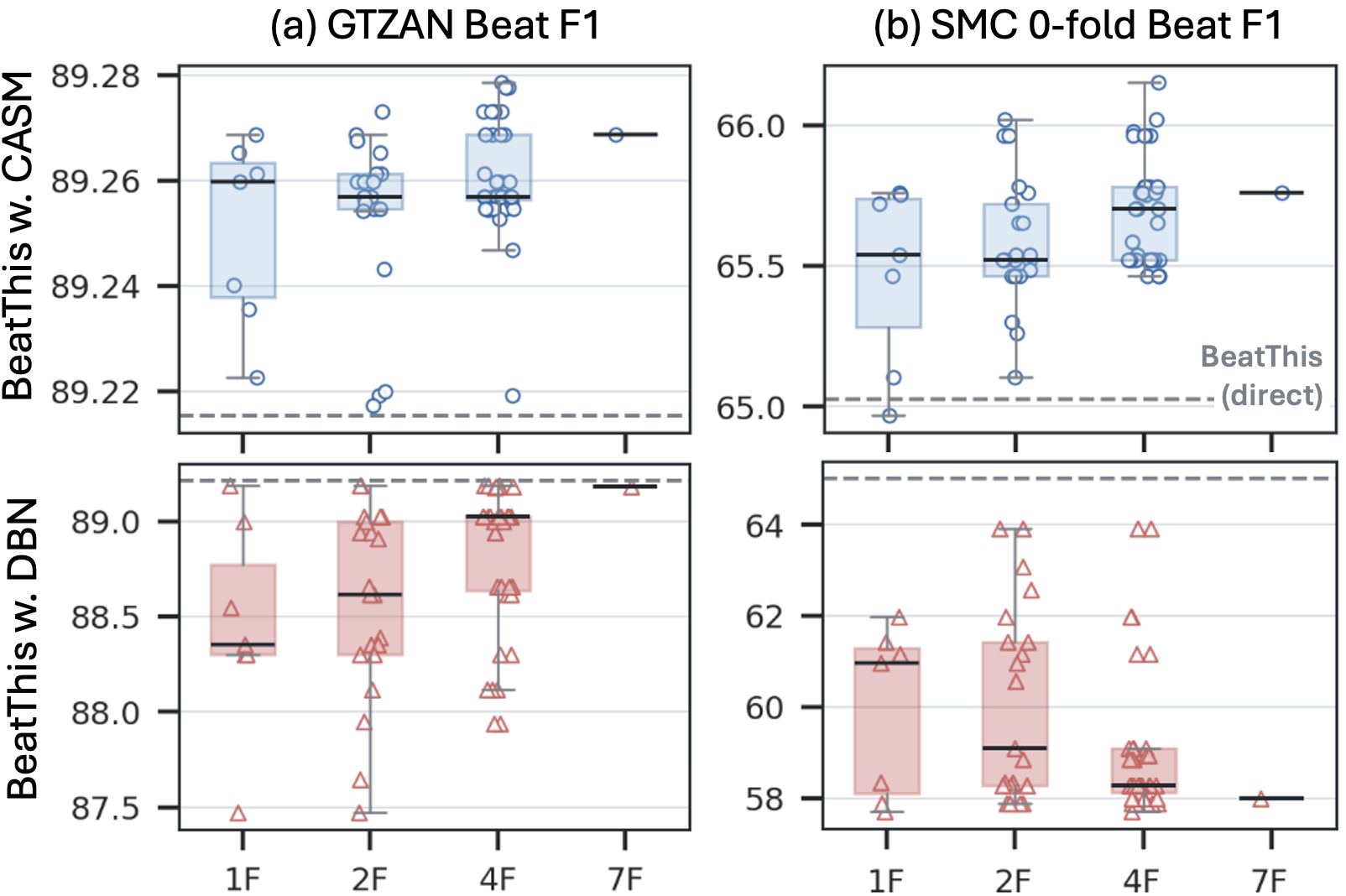}
\caption{Sensitivity to calibration data for CASM and DBN. Each point is one decoder configuration, calibrated on one of the 7/21/35 possible 1-/2-/4-fold subsets of SMC folds 1--7 (7F: all seven folds, a single configuration) and evaluated on SMC fold 0 and GTZAN. Because the subsets overlap, the boxes describe spread only. Dashed lines show the direct peak-picking performance as the reference.}
\label{fig:calibration-scale}
\end{center}
\end{figure}

\vspace{-5pt}

Fig.~\ref{fig:calibration-scale} shows a clear difference in calibration stability. CASM remains tightly clustered across fold combinations and the median remains above Direct at every scale, even when calibrated on a single fold (1F). In contrast, DBN exhibits much wider variation, remains below Direct throughout, and its trends on the two sets diverge: more calibration folds improve GTZAN but degrade SMC. This is consistent with calibration-set over-specialisation in DBN. CASM therefore needs little calibration data, and the 7F setting used in the main tables reflects the full calibration pool rather than a favourable subset.

\section{Conclusion and Future Work}
\label{sec:concl}
We proposed CASM, an offline, model-agnostic post-processor that balances immediate local beat evidence with broader rhythmic patterns. It can strengthen temporal regularity when the local evidence is clear and to rely more on activation evidence when it is ambiguous.  We proved CASM works across different backbones using a single frozen calibration setup. Unlike most post-processors that force a trade-off, CASM uniformly enhances both local beat accuracy (F1) and global continuity (CMLt and AMLt). It is also less sensitive to calibration data than the DBN post-processor, indicating it much easier to tune. Future work can apply CASM to other state-of-the-art beat tracking systems as their code become open-source. Since CASM analyzes neural activations directly, it can ideally work with video, motion, and other multimodal beat tracking systems. Finally, we plan to integrate CASM into neural network training loops, similar to how PLPDP evolved into dPLP \cite{chiu2025dplp} or how SemiCRF is used in piano transcription \cite{yan2024semicrf}.

\section{Acknowledgments.}
The computational resources are supported by SongShan Lake HPC Center (SSL-HPC) at Great Bay University. This work was conducted while the first and second author were visiting student researchers at Great Bay University.

\bibliographystyle{IEEEbib}
\bibliography{casm}

\end{document}